\documentclass[
11pt,
reprint,
prl,
superscriptaddress,
longbibliography,
]{revtex4-2}

\usepackage{graphicx}
\usepackage{dcolumn}
\usepackage{bm}
\usepackage{hyperref}
\usepackage[mathlines]{lineno}
\usepackage{braket}
\usepackage{blkarray, dsfont}
\usepackage{amsmath}

\begin{document}

\title{High-fidelity gates with mid-circuit erasure conversion in a metastable neutral atom qubit}

\author{Shuo Ma}
\thanks{These authors contributed equally to this work.}
\affiliation{Department of Electrical and Computer Engineering, Princeton University, Princeton, NJ 08544, USA}
\affiliation{Department of Physics, Princeton University, Princeton, NJ 08544, USA}

\author{Genyue Liu}
\thanks{These authors contributed equally to this work.}
\affiliation{Department of Electrical and Computer Engineering, Princeton University, Princeton, NJ 08544, USA}

\author{Pai Peng}
\thanks{These authors contributed equally to this work.}
\affiliation{Department of Electrical and Computer Engineering, Princeton University, Princeton, NJ 08544, USA}

\author{Bichen Zhang}
\affiliation{Department of Electrical and Computer Engineering, Princeton University, Princeton, NJ 08544, USA}

\author{Sven Jandura}
\affiliation{University of Strasbourg and CNRS, CESQ and ISIS (UMR 7006), aQCess, 67000 Strasbourg, France}

\author{Jahan Claes}
\affiliation{Department of Applied Physics, Yale University, New Haven, CT 06520, USA}
\affiliation{Yale Quantum Institute, Yale University, New Haven, CT 06511, USA}

\author{Alex P. Burgers}
\thanks{Current affiliation: Department of Electrical and Computer Engineering, College of Engineering, University of Michigan, Ann Arbor, MI 48109, USA.}
\affiliation{Department of Electrical and Computer Engineering, Princeton University, Princeton, NJ 08544, USA}

\author{Guido Pupillo}
\affiliation{University of Strasbourg and CNRS, CESQ and ISIS (UMR 7006), aQCess, 67000 Strasbourg, France}

\author{Shruti Puri}
\affiliation{Department of Applied Physics, Yale University, New Haven, CT 06520, USA}
\affiliation{Yale Quantum Institute, Yale University, New Haven, CT 06511, USA}

\author{Jeff D. Thompson}
\email[ ]{jdthompson@princeton.edu}
\affiliation{Department of Electrical and Computer Engineering, Princeton University, Princeton, NJ 08544, USA}

\begin{abstract}
The development of scalable, high-fidelity qubits is a key challenge in quantum information science. Neutral atom qubits have progressed rapidly in recent years, demonstrating programmable processors~\cite{bluvstein2022,graham2022a} and quantum simulators with scaling to hundreds of atoms~\cite{ebadi2021,scholl2021}. Exploring new atomic species, such as alkaline earth atoms~\cite{cooper2018,Norcia2018,Saskin2019}, or combining multiple species~\cite{singh2022} can provide new paths to improving coherence, control and scalability. For example, for eventual application in quantum error correction, it is advantageous to realize qubits with structured error models, such as biased Pauli errors~\cite{bonillaataides2021} or conversion of errors into detectable erasures~\cite{wu2022}. In this work, we demonstrate a new neutral atom qubit, using the nuclear spin of a long-lived metastable state in ${}^{171}$Yb. The long coherence time and fast excitation to the Rydberg state allow one- and two-qubit gates with fidelities of 0.9990(1) and 0.980(1), respectively. Importantly, a significant fraction of all gate errors result in decays out of the qubit subspace, to the ground state. By performing fast, mid-circuit detection of these errors, we convert them into erasure errors; during detection, the induced error probability on qubits remaining in the computational space is less than $10^{-5}$. This work establishes metastable ${}^{171}$Yb as a promising platform for realizing fault-tolerant quantum computing.
\end{abstract}

\maketitle
\onecolumngrid

Neutral atoms in optical tweezer arrays are a rapidly developing field for quantum science~\cite{kaufman2021}, including programmable quantum processors~\cite{bluvstein2022,graham2022a} and many-body simulators~\cite{browaeys2020}. Recent advances include scaling to hundreds of atoms~\cite{scholl2021,ebadi2021}, dual-species arrays with mid-circuit measurements and continuous reloading~\cite{singh2022,singh2022a}, and efficient architectures for quantum error correction~\cite{cong2022,wu2022}. The development of tweezer arrays using alkaline earth atoms~\cite{cooper2018,Norcia2018,Saskin2019} has also led to applications to atomic clocks~\cite{norcia2019, madjarov2019}, and  long-lived nuclear spin qubits~\cite{barnes2022,ma2022,jenkins2022}.

An important feature of alkaline earth atoms is a metastable excited electronic state, which can be used to encode information instead of (or in addition to) the ground state~\cite{wu2022,chen2022}. This creates a number of unique possibilities including fast, high-fidelity excitation to the Rydberg state~\cite{madjarov2020} and mid-circuit fluorescence measurements or laser cooling of ground state atoms. The latter feature is important for mid-circuit readout and qubit reloading, playing a role analogous to a second atomic species. Metastable qubits have been proposed for both neutral atoms~\cite{wu2022,chen2022} and ions~\cite{allcock2021,campbell2020}, and recently demonstrated with $^{171}$Yb$^+$ ions~\cite{yang2022,roman2021}.

The metastable qubit encoding also enables mid-circuit detection of errors resulting in transitions to the ground state~\cite{campbell2020,wu2022}. This converts these errors into erasure errors~\cite{grassl1997,bennett1997}, which are significantly easier to correct in the context of fault-tolerant quantum computing~\cite{barrett2010,wu2022,sahay2023}. If a large fraction of all errors are converted into erasures, and the information about which qubits were erased can be extracted while preserving the quantum state of qubits that did not have errors, the resource overhead for fault-tolerant computing is significantly reduced. This concept has stimulated new qubit designs in several platforms~\cite{wu2022,kubica2022,kang2022,teoh2022,tsunoda2022,lu2023}, but has not been experimentally demonstrated.

In this work, we demonstrate a qubit encoded in the nuclear spin of the metastable $6s6p\,^3P_0$ state in neutral $^{171}$Yb. We demonstrate seconds-scale lifetimes and coherence times, single-qubit gates with $\mathcal{F} = 0.9990(1)$, and two-qubit gates with $\mathcal{F} = 0.980(1)$, where the latter is enabled by a novel gate design~\cite{jandura2022} and fast, single-photon excitation to the Rydberg state. A significant fraction of the gate errors result in transitions out of the metastable state, to the atomic ground state. By performing fast ($20\,\mu$s) imaging of ground state atoms~\cite{bergschneider2018}, we are able to detect these leakage errors mid-circuit, converting them into erasure errors, with a probability less than $10^{-5}$ of inducing an error on qubits remaining in the metastable state during detection. We show that 56\% of single-qubit gate errors and 33\% of two-qubit gate errors are detected in this manner. We conclude by discussing future opportunities for the metastable $^{171}$Yb qubit including  improvements in the gate fidelity and erasure fraction, and mid-circuit qubit readout.

Our experiment begins by trapping individual $^{171}$Yb atoms in an array of optical tweezers (Fig. 1a)~\cite{ma2022,jenkins2022}. We use optical pumping to initialize the qubit in the state $\ket{1} \equiv \ket{6s6p\,^3P_0, F=1/2, m_F = +1/2}$ (Fig. 1b). Single-qubit rotations are driven using an RF magnetic field tuned to the nuclear spin Larmor frequency $\omega_L = 2\pi \times 5.70$ kHz ($|B| = 5.0$ G, Fig. 1c). Spin readout is implemented by removing atoms in $\ket{1}$ from the trap (via excitation to the Rydberg state and subsequent autoionization~\cite{ma2022}), then depumping the remaining metastable population back to the ground state before imaging. The combined fidelity of the state initialization and imaging is 0.981(9), limited by loss during the depumping step.

The absence of hyperfine coupling in this state allows for extremely long coherence times for the nuclear spin ($T_1 = 23(14)$\,s, $T_2^* = 0.92(2)$s), as demonstrated previously for nuclear spin qubits in the ground state~\cite{barnes2022,ma2022,jenkins2022}. Two-qubit operations are performed by selectively exciting the state $\ket{1}$ to the Rydberg state $\ket{6s59s\,^3S_1\,F=3/2, m_F=3/2}$ using a 302 nm laser. In contrast to the metastable state, the presence of hyperfine coupling in the Rydberg manifold results in a large Zeeman shift between magnetic sublevels in the Rydberg state, such that two-qubit operations can be performed significantly faster than $\omega_L^{-1}$  (Fig. 1d). The same concept can be used to implement fast single-qubit rotations with a lower-lying excited state~\cite{jenkins2022}.

The fidelity of idling and single-qubit operations is limited primarily by the finite lifetime of the metastable state (Fig. 2a), which is $\Gamma_m^{-1} = 2.96(12)$ s under typical operating conditions. The decay rate depends strongly on the trap power, and can be described as $\Gamma_m = \Gamma_0 + \alpha P + \beta P^2$ (Fig. 2b). The constant term $\Gamma_0 = 0.4(2)$\,s$^{-1}$ includes background loss and radiative decay (0.05\,s$^{-1}$). The linear [$\alpha = 0.20(7)$ s$^{-1}$mW$^{-1}$] and quadratic [$\beta = 0.053(5)$ s$^{-1}$mW$^{-2}$] terms are attributed to Raman scattering and photoionization, respectively, where the latter is possible because the trapping laser energy ($\lambda = 486.8$ nm) is above the two-photon ionization limit. However, both radiative decay and Raman scattering return the atom to the ground state (with suitable repumping of the other metastable state, $^3P_2$), enabling eventual detection as shown in Fig. 2c.

To detect decays to the ground state without disturbing qubits in the metastable state, we use fast fluorescence imaging~\cite{bergschneider2018} on the strong $^1S_0 - {}^1P_1$ transition. This transition is approximately 160 times faster than the $^1S_0$ - $^3P_1$ intercombination transition used to initialize the atom array and measure the final spin state~\cite{Saskin2019}, which allows for shorter acquisition times at the expense of losing the atom after the image. This tradeoff is favorable when probing for qubits that have already had errors, to minimize the probability of additional decays during the imaging time. By illuminating the array with counter-propagating beams near saturation, we can detect atoms in the ground state after a 20\,$\mu$s exposure, with a fidelity of 0.986 (Fig. 2d).

We now apply this technique to demonstrate mid-circuit detection of decay errors, converting them into erasures. We use a standard randomized benchmarking (RB) circuit with up to 300 single-qubit gates, and acquire a fast image after every 50 gates to probe for atoms that have decayed to the ground state (Fig. 2e). The average gate error rate is $\epsilon = 1.0(1) \times 10^{-3}$. However, by conditioning on the absence of a ground state atom in all of the erasure detection steps before the final qubit measurement (Fig. 2f), the error rate decreases to $\epsilon_{c} = 4.5(3) \times 10^{-4}$. Therefore, 56(4)\% of the errors are detected mid-circuit and converted into erasure errors. Some of the errors that are not converted can be attributed to undetected loss (\emph{i.e.}, background loss and photoionization, $2 \times 10^{-4}$) and scattering back into the metastable state while repumping $^3P_2$ on a non-ideal transition ($1.0(6) \times 10^{-4}$). The remainder are errors within the metastable state ($\approx 1.5 \times 10^{-4}$).

We now consider imperfections in the erasure detection process, which we divide into two types: errors induced on the qubits remaining in the metastable state, and imperfect detection of atoms in the ground state. Qubits remaining in the metastable state suffer errors from the additional decay probability during the imaging time, $P_d = 7 \times 10^{-6}$, or from the off-resonant scattering of the imaging light. The latter effect is strongly suppressed by the large detuning between the $^1S_0 - {}^1P_1$ transition and any transitions originating from the metastable state (the nearest state is detuned by $2\pi \times 22$\,THz). We probe for scattering errors by continuously illuminating the atoms with the imaging light during the RB sequence (Fig. 2e, red star). No effect is observed, bounding this error at $< 10^{-6}$ per imaging time. In Fig. 2g, we examine the image fidelity by varying the erasure detection threshold and tracking two quantities: the probability $P(\mathrm{err.|det.})$ that a qubit has an error at the end of the circuit, given that a detection event occurred, and the fraction of all errors that are detected prior to the end of the circuit, $R_e = (\epsilon - \epsilon_c)/\epsilon$. The first quantity is ideally 1 and decreases with false positive detections when the threshold is too low. The second quantity is also ideally 1 but is limited by the fraction of errors that are not detectable, as well as false negative detections. We find that a suitable threshold exists where essentially all detectable errors are detected, but the false positive rate remains small.

We now turn to two-qubit entangling gates. We implement a controlled-Z (CZ) Rydberg blockade gate using the time optimal gate protocol of Ref.~\cite{jandura2022}, which is a continuous-pulse gate based on the symmetric controlled-Z (CZ) gate of Ref.~\cite{Levine2019}. The specific gate used in this work is further optimized to compensate for off-resonant transitions between both qubit states and other Rydberg levels (Fig. 3a,b). Precise control over the Rydberg laser pulse is achieved by coupling the laser into a UV-stable fiber~\cite{colombe2014}, and monitoring the transmitted pulse using a heterodyne receiver. A UV power of 6 mW is incident on the atoms, corresponding to a Rabi frequency $\Omega_{UV} = 2\pi \times 1.6$\,MHz.

To demonstrate the basic functionality of the gate, we prepare and measure a Bell state in parallel on five pairs of atoms (Fig. 3c). We obtain a raw Bell state fidelity of $\mathcal{F}=0.866(12)$, and estimate an intrinsic fidelity for the entanglement step of $\mathcal{F}=0.99(2)$ by separately characterizing state preparation and measurement (SPAM) errors. We use a different measurement circuit than previous works~\cite{Levine2019,ma2022}, which increases certain SPAM errors but makes them easier to characterize. In our approach, we always record the fraction of events in the state $\ket{00}$, where both atoms are bright, using single-qubit rotations to map other desired observables onto this state. We then characterize the SPAM error by running the same sequence with the CZ gate removed, finding a SPAM fidelity of $\mathcal{F}_{sp} = 0.872(6)$. The intrinsic Bell state creation fidelity is estimated by renormalizing all measurement outcomes by $\mathcal{F}_{sp}$.

To more precisely characterize the performance of the CZ gate, we perform a randomized-benchmarking-type experiment with up to 10 CZ gates interspersed with random, global single-qubit rotations. We find an error probability of $2.0(1) \times 10^{-2}$ per gate, corresponding to a fidelity of 0.980(1). We note that using global single qubit gates invalidates the rigorous guarantees of two-qubit randomized benchmarking, and is insensitive to certain types of errors~\cite{baldwin2020}. However, we have simulated this benchmarking approach using a realistic model of the atomic dynamics over a wide range of error rates, and find that it gives a consistent lower bound to the true fidelity (see Methods). From a detailed model with independently measured parameters, we infer that the leading sources of gate error are the finite lifetime of the Rydberg state ($65(2)\,\mu$s, $4 \times 10^{-3}$ error) and Doppler shifts ($T=2.9\,\mu$K, $5 \times 10^{-3}$ error).

A large fraction of these errors results in leakage outside of the qubit space, through spontaneous decay from $\ket{r}$ to low-lying states, or as population remaining in $\ket{r}$ or other Rydberg states (populated via black-body radiation) at the end of the gate. Leakage errors are intrinsic to Rydberg gates in any atomic species and typically result in undetected loss~\cite{cong2022}. By taking advantage of the unique property of alkaline earth atoms that the Rydberg states remain trapped in the optical tweezer by the Yb$^+$ ion core polarizability~\cite{wilson2022}, we can recapture and detect this leaked population by waiting for it to decay. In Fig. 4b, we show that for an atom initially prepared in $\ket{r}$, we recover 10\% of the population in $^3P_0$, 25\% in $^1S_0$, and 35\% in $^3P_2$ after 400$\,\mu$s (approximately 30\% of the decays are unaccounted for). After repumping $^3P_2$ via $^3S_1$, 51\% of the population is in $^1S_0$, and 19\% in $^3P_0$.

To convert two-qubit gate errors into erasures, we run the benchmarking circuit with interleaved fast imaging (after every two CZ gates). We find a lower error rate after conditioning on not detecting a ground state atom, $\epsilon_c = 1.3(1) \times 10^{-2}$ per gate (Fig. 4c). This corresponds to converting approximately 33\% of the errors into erasure errors. Our error model predicts that 60\% of all gate errors are leakage, which is consistent with the experiment given that only half of the Rydberg leakage is detected (see Methods). Waiting for the Rydberg population to decay increases the erasure detection time from $20\,\mu$s to $420\,\mu$s, increasing the decay probability on qubits without errors to $P_d = 1.4 \times 10^{-4}$.

Finally, we demonstrate that erasure errors occur asymmetrically from the qubit states $\ket{0}$ and $\ket{1}$, which is another form of bias that is advantageous in the design of fault-tolerant systems~\cite{sahay2023}. In Fig. 4e, we show the probability per gate of detecting an erasure when preparing in $\ket{00}$, $\ket{++}$, and $\ket{11}$, using a longer sequence with up to 18 CZ gates (with no interleaved single-qubit rotations, to maintain the state populations in the computational basis). The probability to detect a leaked qubit is much higher when the initial state has a probability to be in $\ket{1}$, as expected from the selective excitation of $\ket{1}$ to the Rydberg state~\cite{cong2022}. We infer a lower bound on the ratio of the erasure probabilities of $p_{11}/p_{00} > 15(9)$.

Having demonstrated the basic properties of the metastable qubit and erasure conversion, we now reconsider the advantages and disadvantages of metastable qubit encoding. The main disadvantage is that the finite metastable state lifetime introduces an additional error channel that affects very slow operations and qubit storage. The lifetime is 2-3 times shorter than typical Raman-limited $T_1$ times for hyperfine qubits in alkali atoms~\cite{bluvstein2022,graham2022a}, but the impact is offset by the fact that many of these decays can be detected, as demonstrated here. The metastable state decay is not relevant on the sub-microsecond timescale of two-qubit gates, which are instead limited by the finite decay rate of the Rydberg state. The same decay channel is present for Rydberg gates in any atomic species and typically results in atom loss. The fact that the metastable qubit allows these decays to be recaptured and detected is a significant advantage, which will become even more significant as other sources of error are eliminated and Rydberg decays become dominant~\cite{wu2022,jandura2022a}. Using fast single-photon excitation from the metastable state, as demonstrated here, our error model predicts that significant improvements in gate fidelity are achievable with modest laser upgrades (see Methods).

There are several straightforward improvements to increase the fraction of errors that can be detected and converted into erasures. The fraction of detected metastable state decays can be increased by using an alternative repumping transition for $^3P_2$ (see Methods), a longer trapping wavelength to suppress photoionization, and better vacuum to reduce background losses. There is no apparent obstacle to detecting virtually all of the metastable state decays. In two-qubit gates, the fraction of all errors that are leakage will increase as the gate fidelity improves and Rydberg decays become the dominant error mechanism, and could be as high as 98\%~\cite{wu2022}. Composite pulses can also convert certain errors such as amplitude noise and Doppler shifts into erasures~\cite{jandura2022a,fromonteil2022}. Finally, the detection fidelity of Rydberg leakage can be increased by identifying and repumping additional decay pathways, or by ionizing atoms in the Rydberg state at the end of the gate and detecting the Yb$^+$ ions directly using fluorescence~\cite{wu2022,mcquillen2013} or charged particle detectors, which have demonstrated 98\% ion detection efficiency from an optical tweezer~\cite{henkel2010}.

The demonstrated high-fidelity control and mid-circuit erasure conversion establishes metastable $^{171}$Yb as a promising architecture for fault-tolerant quantum computing based on erasure conversion. In this context, the key property of this demonstration is that the quantum state of qubits that did not have erasures is unaffected by the erasure detection process. In an error correcting code, this will allow the erased qubits to be replaced using a moveable optical tweezer, such that the correct code state can be restored after measuring the error syndromes~\cite{wu2022}.

The detection protocols demonstrated in this work can also be adapted to mid-circuit qubit readout~\cite{singh2022a,deist2022,graham2023}, by selectively transferring one of the qubit levels back to the ground state (\emph{i.e.}, using optical pumping). Subsequently transferring the other level and measuring the ground state population again would also allow atom loss to be distinguished, which is also beneficial for error correction~\cite{sahay2023}. The metastable qubit will also enable mid-circuit reloading of new qubits, as demonstrated already with dual-species experiments~\cite{singh2022a}.

\emph{Acknowledgements} We acknowledge helpful conversations with Shimon Kolkowitz and Michael Gullans. This work was supported by the Army Research Office (W911NF-1810215), the Office of Naval Research (N00014-20-1-2426), DARPA ONISQ (W911NF-20-10021), the National Science Foundation (QLCI grant OMA-2120757), and the Sloan Foundation. This research also received funding from the European Union’s Horizon 2020 program under the Marie Sklodowska-Curie project 955479 (MOQS), the Horizon Europe program HORIZON-CL4-2021- DIGITAL-EMERGING-01-30 via the project 101070144 (EuRyQa) and from the French National Research Agency under the Investments of the Future Program project ANR-21-ESRE-0032 (aQCess).

\emph{Competing interests} G.P. is co-founder and shareholder of QPerfect.

\emph{Note:} While finalizing this manuscript, we became aware of recent, related work in Refs.~\cite{evered2023,scholl2023}.

\clearpage

\begin{figure*}[htp!]
    \centering
    \includegraphics[width=180 mm]{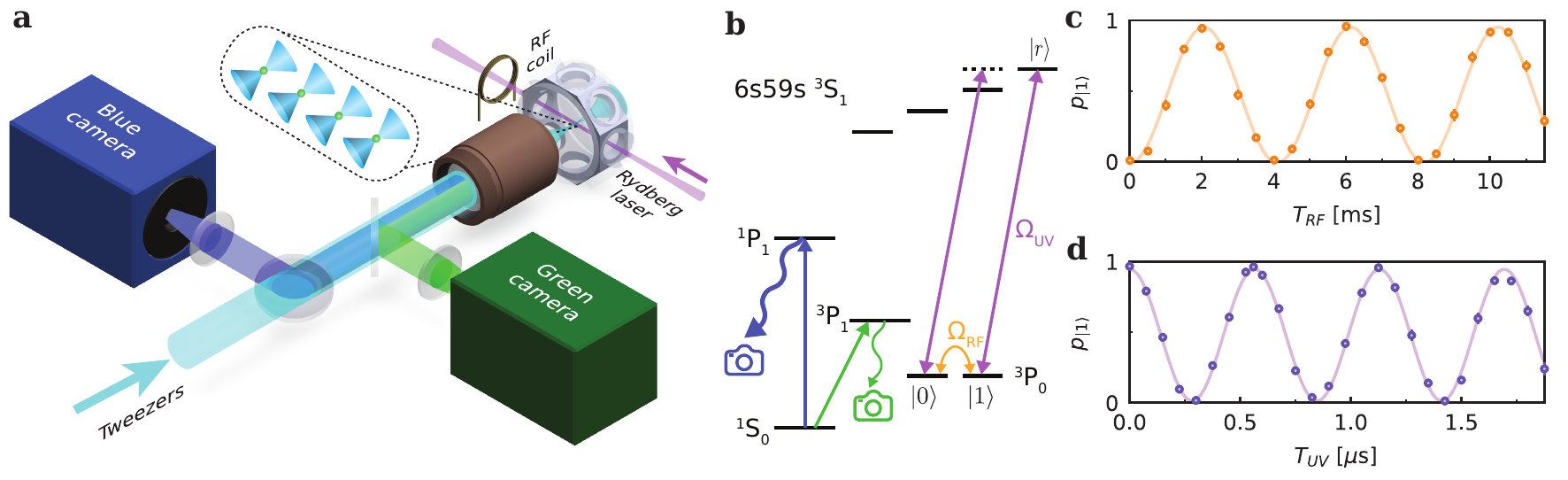}
    \caption{\textbf{Metastable $^{171}$Yb qubit} (a) Overview of the experimental apparatus, showing an array of optical tweezers inside a glass cell, and cameras for acquiring nondestructive images (using the $^1S_0-{}^3P_1$ transition, 556 nm, $\Gamma = 2\pi \times 182$\,kHz) and fast images (using the $^1S_0-{}^1P_1$ transition, 399 nm, $\Gamma = 2\pi \times 29$\,MHz). (b) Abbreviated $^{171}$Yb level diagram showing the ground state, imaging transitions, and the nuclear spin sublevels within the metastable $^3P_0$ state that encode the qubit. Single-qubit gates are generated with an RF magnetic field $\Omega_{RF}$, and entangling gates are implemented by coupling $\ket{1}$ to a Rydberg state, $\ket{r}$, with an ultraviolet (UV) laser at 302 nm. (c) Nuclear spin Rabi oscillation between $\ket{0}$ and $\ket{1}$, with a $\pi$-pulse time of 2.0 ms. (d) Rabi oscillation between $\ket{1}$ and $\ket{r}$, with a $\pi$-pulse time of 330 ns.}
    \label{fig:intro}
\end{figure*}

\begin{figure*}[htp!]
    \centering
    \includegraphics[width=180 mm]{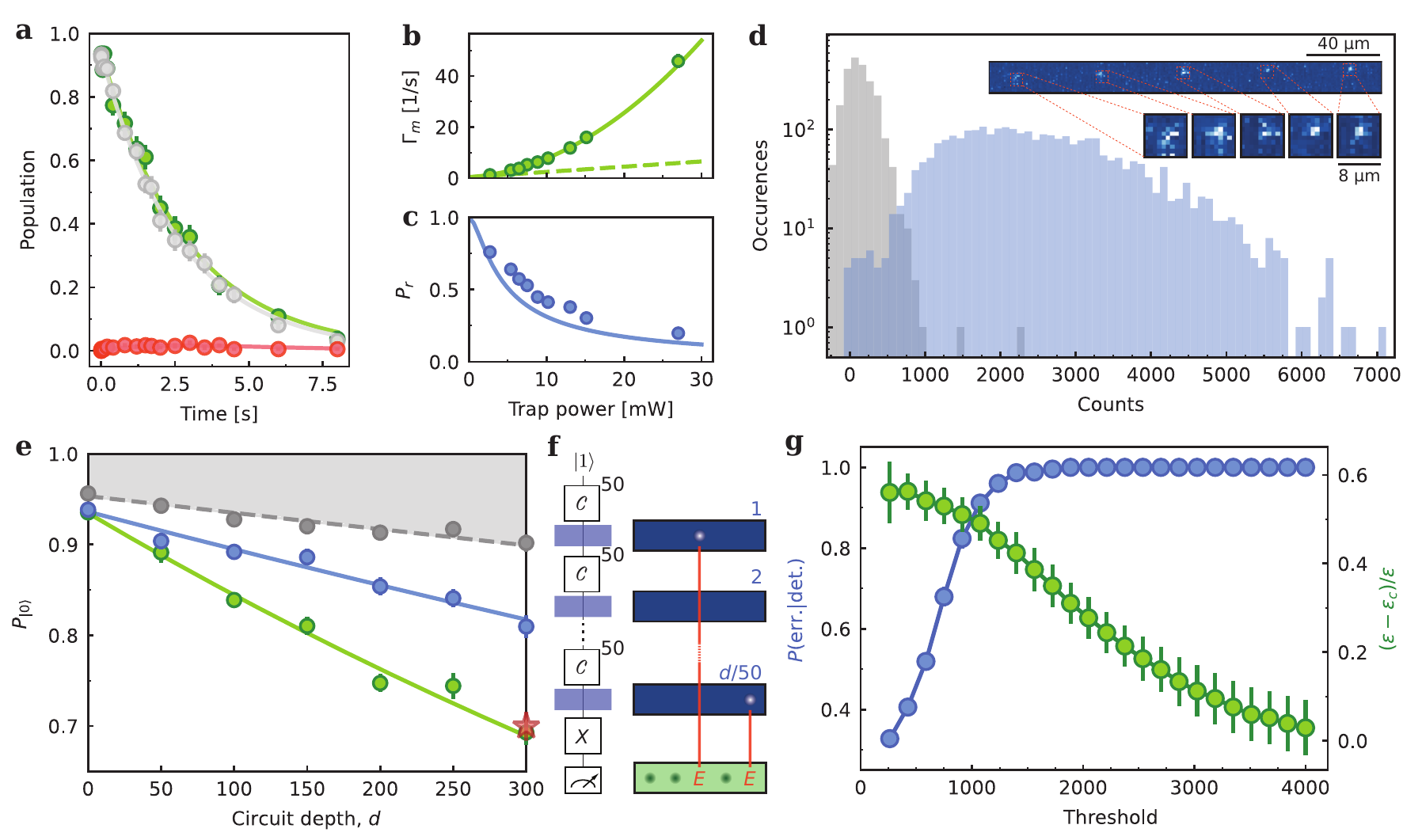}
    \caption{\textbf{Single-qubit gates with mid-circuit erasure conversion} (a) Lifetime of the $^{171}$Yb metastable qubit in an optical tweezer (power $P=0.76$ mW, depth $U = 58\,\mu$K). The green points show the total metastable state population ($1/e$ decay time $\Gamma_m^{-1} = 2.96(12)$\,s), while the gray and red points show the population $P_0$ in $\ket{0}$, after initializing in $\ket{0}$ and $\ket{1}$, respectively. Fitting the $\ket{0}$ and $\ket{1}$ population yields an average spin-flip time $T_1 = 23(14)$ s. (b) Metastable state decay rate $\Gamma_m$ as a function of trap power, showing the quadratic model (see text) and its linear part (dashed line). (c) Probability $P_r$ to recover an atom in the ground state after a decay from the metastable state. (d) Histogram of camera counts from fast ($20\,\mu$s) images on the $^1S_0 - {}^1P_1$ transition. The discrimination fidelity is 0.986. \emph{Inset:} example single-shot fast image of a 5-site array. (e,f) Randomized benchmarking (RB) of single-qubit gates, using the circuit shown in panel f). After every 50 Clifford gates ($\mathcal{C}$), a fast image probes population in the ground state, converting the decay into an erasure error, $E$. The total error rate is $\epsilon = 1.0(1)\times 10^{-3}$ (green), which falls to $\epsilon_c = 4.5(3)\times 10^{-4}$ after conditioning on not detecting a ground state atom before the end of the circuit (blue). The total atom survival probability is shown in grey. The red star is from a second experiment with the fast imaging light left on continuously, showing no change. (g) The threshold for detecting a ground state atom in the analysis of the fast images affects the erasure conversion performance. We quantify this using the probability to have an error at the end of the RB sequence conditioned on detecting an erasure, $P(\mathrm{err.|det.})$ (blue), and the fraction of all errors that are detected as erasures, $(\epsilon-\epsilon_c)/\epsilon$ (green). A threshold near 700 is used in the analysis in panel e).}
    \label{fig:1qgate}
\end{figure*}

\begin{figure}[htp!]
    \centering
    \includegraphics[width=90 mm]{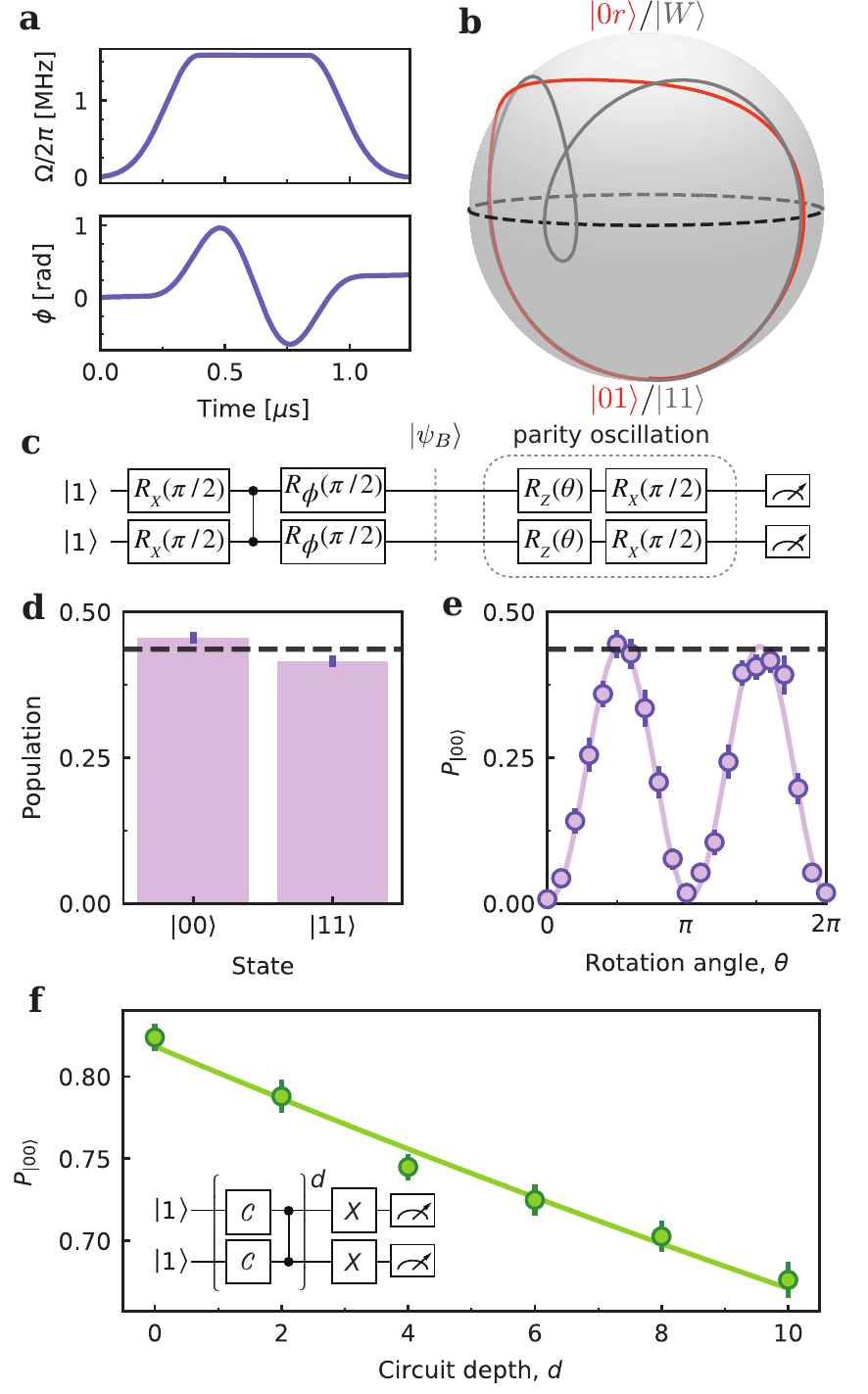}
    \caption{\textbf{Time-optimal two-qubit gates} (a) Laser amplitude (top) and phase (bottom) for the time-optimal CZ gate~\cite{jandura2022}. (b) Bloch sphere trajectories during the gate in the $\{\ket{01},\ket{0r}\}$ subspace (red) and $\{\ket{11},\ket{W}\}$ subspace (grey). Here, $\ket{W}\equiv(\ket{1r}+\ket{r1})/\sqrt{2}$. (c) Gate sequence used to prepare and characterize the Bell state $\ket{\psi_B} = (\ket{00} + i\ket{11})/\sqrt{2}$. (d) Population of $\ket{00}$ and $\ket{11}$ in the Bell state. The dashed line shows the probability to prepare and measure the bright state $\ket{00}$ without the CZ gate. (e) Parity oscillations showing the coherence of the Bell state, measured using only the bright state population. The off-diagonal part of the Bell state density matrix is $P_c=4A$, where $A$ is the $\cos(2\theta)$ oscillation amplitude. The Bell state fidelity is $(P_{00}+P_{11}+P_c)/2$. (f) Randomized circuit characterization of the two-qubit gate, with an error $\epsilon = 2.0(1) \times 10^{-2}$ per gate.}
    \label{fig:bellstate}
\end{figure}

\begin{figure}[htp!]
    \centering
    \includegraphics[width=90 mm]{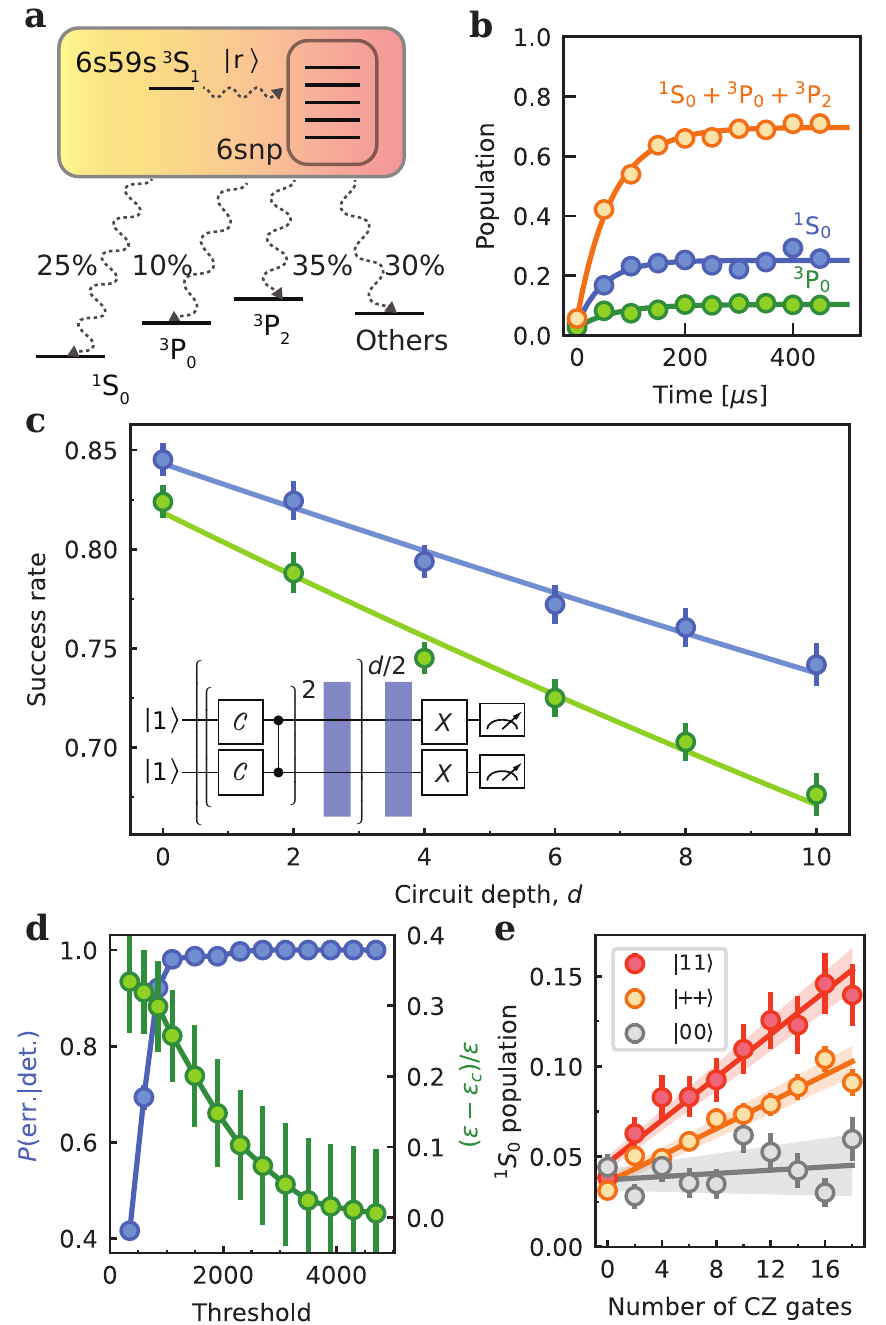}
    \caption{\textbf{Erasure conversion for two-qubit gates} (a,b) Population leaked into $\ket{r}$ or other Rydberg states can be recovered by waiting for it to decay. (c) Randomized circuit characterization of CZ gates with interleaved erasure detection after every two gates. The total error rate of $\epsilon = 2.0(1) \times 10^{-2}$ per gate (green) is reduced to $\epsilon_c = 1.3(1) \times 10^{-2}$ per gate (blue) after conditioning on not detecting an atom in the ground state before the end of the circuit. We note that the green curve is reproduced in Fig. 3f. (d) Analysis of the erasure detection fidelity during two-qubit gates, following Fig.~2g. (e) Erasure detection probability for different initial states under repetitive CZ gates. Linear fits are overlaid, with the shaded area marking one standard deviation. The erasure probability for $\ket{00}$ is $4(6)\times 10^{-4}$ per gate, consistent with zero.}
    \label{fig:srb}
\end{figure}

\clearpage

\section{\label{sec:methods} Methods}

\renewcommand{\thefigure}{S\arabic{figure}}
\setcounter{figure}{0}

\subsection{Experimental apparatus}

We load a tweezer array from a 3D magneto-optical trap operating on the $^1S_0 \rightarrow {}^3P_1$ intercombination line. The tweezers are at a wavelength of $\lambda$ = 486.78 nm, which is magic for the ground state and the $\ket{^3P_1, F=3/2, |m_F|=1/2}$ excited states~\cite{ma2022}. An acousto-optic deflector driven by an arbitrary waveform generator is used to create defect-free 1D tweezer arrays via rearrangement~\cite{endres2016,barredo2016}. We use approximately 4 mW/tweezer in the plane of the atoms for loading and rearrangement, corresponding to a trap depth of 300 $\mu$K. After loading, a short blue-detuned cooling pulse is used to reach a temperature of approximately $T=5\,\mu$K.
For determining the initial and final tweezer occupation, we image using the intercombination line transition, achieving a fidelity and survival probability of 0.995 in a 15~ms exposure time~\cite{Saskin2019}. The images are acquired with an sCMOS camera (Photometrics Prime BSI). To perform fast imaging~\cite{bergschneider2018} for detecting atoms that have decayed to $^1S_0$, we use the $^1S_0 \rightarrow {}^1P_1$ transition at 399 nm. We illuminate the atoms with a resonant retro-reflected beam at a power of $I = 4 I_{sat}$. The image is acquired using an EMCCD camera (N\"uv\"u HN\"u 512 Gamma) with EM gain = 1000. We achieve a detection fidelity of 0.986 in 20\,$\mu$s. The survival probability of this imaging process is small but non-zero (2-5\%), so we follow each blue image with an additional 80\,$\mu$s pulse to ensure that none of the atoms that decayed to the ground state are present in the final spin measurement image. The position spread of the atoms during the imaging is approximately 2\,$\mu$m (r.m.s).

We can create arrays of up to 30 optical tweezers, limited by the available laser power. To avoid Rydberg interactions during the spin readout, we use a spacing of $d= 43\,\mu$m for the experiments in Fig. \ref{fig:1qgate}d, limiting the array to 5 sites. For the two-qubit experiments in Figs.~\ref{fig:srb} and \ref{fig:bellstate}, we use five dimers spaced by $d=43\,\mu$m, with a separation of 2.4\,$\mu$m between the atoms in each pair.

\begin{figure*}
    \centering
    \includegraphics[width=125mm]{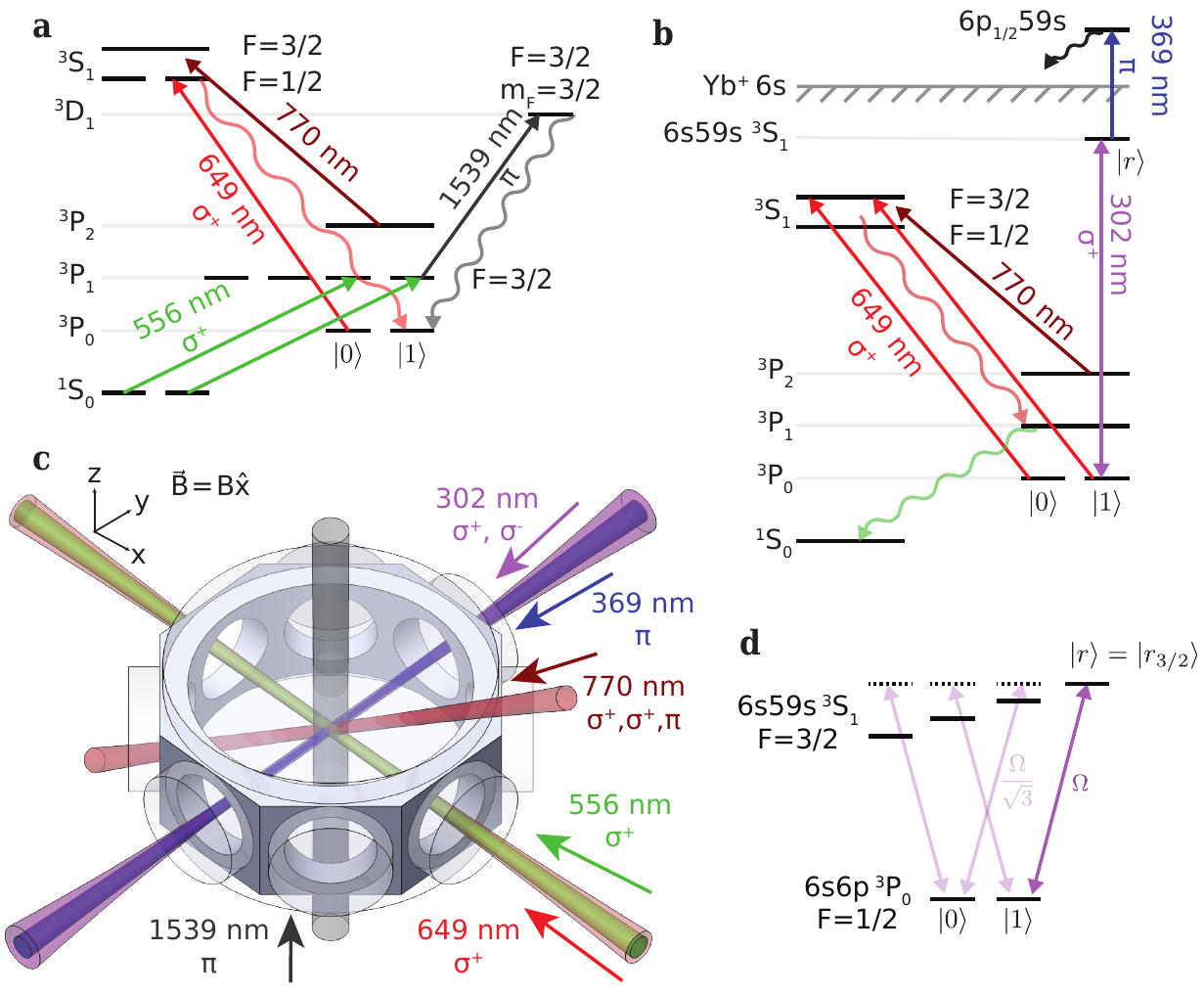}
    \caption{\textbf{Level diagrams and laser beam geometry} (a) Partial level diagram showing the transitions used to optically pump into the state $\ket{1}$ for initialization. (b) Partial level diagram indicating the transitions used to measure the spin state in $^3P_0$. First, atoms in $\ket{1}$ are removed from the trap using Rydberg excitation and subsequent autoionization. Then, all population in $^3P_0$ is pumped back to $^1S_0$ and imaged. (c) Propagation directions and polarizations of the lasers addressing the atoms. The 556 nm and 399 nm imaging beams are not shown, but are co-propagating with the 770 nm beam and retro-reflected. The microscope objective used to project the tweezers and image the atoms (numerical aperture NA=0.6) is positioned above the glass cell. (d) Partial level diagram showing the transitions between $^3P_0$ and the Rydberg manifold used in this work. The detuning between the Rydberg states is 5.8 times larger than $\Omega$. The 302 nm beam is linearly polarized perpendicular to the magnetic field, which is constrained by the geometry of our apparatus. In the future, using a pure $\sigma^+$-polarized 302 nm beam would increase the gate speed by a factor of $\sqrt{2}$ for the same laser power.}
    \label{fig:3P0_init_readout}
\end{figure*}

\subsection{Metastable state initialization and measurement}
We initialize atoms into the $^3P_0$ state using optical pumping, to avoid the need for a clock laser and state-insensitive tweezers for the clock transition. The optical pumping scheme is depicted in Fig.~\ref{fig:3P0_init_readout}a. A coherent two-photon transition (556 nm, $\sigma^+$-polarized; 1539 nm, $\pi$-polarized) is used to excite atoms from $\ket{^1S_0, m_F = +1/2}$ to the $\ket{^3D_1, F=3/2, m_F=+3/2}$ state. The $m_F$ levels in the excited state are split by $5.7$ linewidths in a magnetic field, allowing energy-selective excitation of the $m_F=3/2$ sublevel, even in the presence of polarization imperfections. This state decays to $\ket{1}$ with approximately 64\% probability and returns to $^1S_0$ (via $^3P_1$) in most of the other cases (the branching ratio to $^3P_2$ is 1\%, and this state is continuously repumped via $^3S_1$ using a 770 nm laser with sidebands to address both $F$ levels). The detuning of the 556 nm leg of the two-photon Raman transition is chosen to be resonant with the $\ket{^3P_1, 
F=3/2, m_F=1/2}$ excited state, to continuously pump atoms out of the $\ket{^1S_0, m_F = -1/2}$ state.

During this process, several percent of the atoms end up in $\ket{0}$, because of off-resonant excitation to other $^3D_1$ states and decays through $^3P_2$. To increase the purity in the $\ket{1}$ state, we apply a short pulse of light at 649 nm ($\sigma^+$), coupling $\ket{0}$ to $\ket{^3S_1, F=1/2,m_F=1/2}$. This removes the atom from $\ket{0}$ with $\approx 90\%$ probability, after which we apply an additional cycle of repumping from the ground state. This process is repeated a second time.

The total duration of the optical pumping process is 500\,$\mu$s, and the average number of scattered photons is less than 2. We measure a temperature of 5.7 $\mu$K for atoms in $^3P_0$, indicating minimal heating.

To measure the population in $^3P_0$, we pump atoms back from $^3P_0$ to the ground state via $^3S_1$ (Fig.~\ref{fig:3P0_init_readout}b), with continuous illumination from 770 nm to repump $^3P_2$. To make this measurement spin-selective, we first remove atoms in $\ket{1}$ by exciting to $\ket{r}$ and autoionizing the Rydberg population~\cite{burgers2022}. In addition to being destructive, this step limits the density of tweezers, to avoid blockade effects. In future work, this step can be replaced with spin-selective optical pumping via $^3S_1$ or $^3D_1$ to allow non-destructive readout. Because the $^3P_2$ state is anti-trapped in the 486 nm tweezer ($U_{{}^3P_2}/U_{{}^3P_0} \approx -2$), we pulse off the traps for 3 $\mu$s during this step, which results in a few percent atom loss probability. This could be mitigated using multiple short modulation pulses or a different tweezer wavelength where $^3P_2$ is trapped (\emph{i.e.}, 532 nm~\cite{Okuno2022}).

We characterize the fidelity of the initialization and readout process by preparing the states $\ket{0}$ and $\ket{1}$ using the procedures described above together with nuclear spin rotations. We observe the correct outcome 99.6(3)\% of the time for $\ket{1}$ (the dark state), and 96.6(8)\% of the time for $\ket{0}$ (the bright state), for an average initialization and readout fidelity of $98.1(9)\%$. We believe the dominant error is loss during the pumping back to $^3P_0$.

Finally, we note several other experimental details. The initialization and readout are performed with a trap depth of $300\,\mu$K, corresponding to a power of 4 mW per tweezer. During the gate operations, the trap is ramped down to $58\,\mu$K (0.76 mW), which reduces Doppler shifts and atom loss from pulsing off the traps during the two-qubit gates. This cools the atoms further to $T=2.94\,\mu$K, measured from the Ramsey coherence time of the Rydberg state.  The 770 nm repumper is left on continuously during the entire time the atom is in the metastable state, to rapidly repump any atoms that scatter or decay to $^3P_2$. This is not the ideal repumper configuration, as it has a 25\% probability to pump an atom back to $^3P_0$. This is evident in the finite spin-flip rate in Fig. 2a --- repeating that measurement without the 770 nm repumper results in no observable spin flips, as expected from the absence of hyperfine coupling~\cite{dorscher2018}. In the future, repumping $^3P_2$ through a $^3D_2$ state (\emph{i.e.}, using transitions at 496.8 nm or 1983 nm) would avoid repopulating $^3P_0$.

\subsection{Rydberg laser system and beam delivery}

\begin{figure*}
    \centering
    \includegraphics[width=150mm]{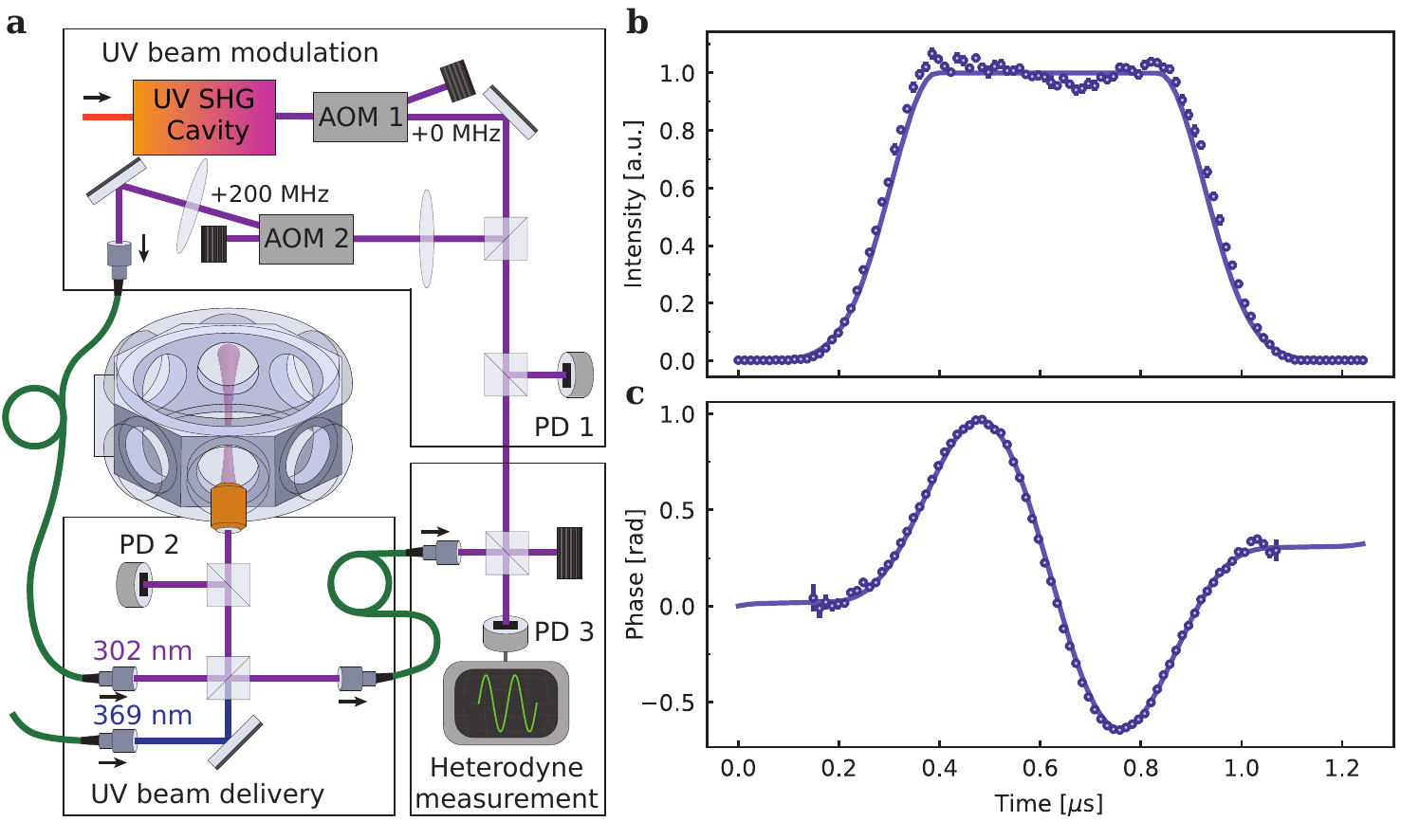}
    \caption{\textbf{Rydberg laser system} (a) The 302 nm light is generated by a resonant cavity. The output beam is power-stabilized by a servo formed by AOM1 and PD1, and pulses are generated by AOM2. The pulsed light is coupled into a fiber and delivered to a monolithic breadboard next to the glass cell. PD2 monitors the pulse power on the breadboard. To monitor the pulse phase, a small fraction of the light is sent back to the optical table with a second fiber, and interfered with the un-modulated laser to form a beatnote on PD3 (at the frequency of AOM2). The beatnote is digitized and digitally demodulated to extract the amplitude and phase profiles shown in (b) and (c), together with the target pulse shapes (solid lines). Programming AOM2 with a naive waveform results in significant phase distortion; the waveform shown in (c) is obtained after closed-loop correction.}
    \label{fig:uv_laser}
\end{figure*}

Atoms are excited to the Rydberg state using UV light at 302 nm. This light is produced in two steps. First, we generate $\approx 1$ W of 604 nm light through sum frequency generation in a 40 mm long PPLN crystal using a titanium-sapphire laser at 980 nm (MSquared Solstis, 1 W) and an Er-doped fiber laser at 1565 nm (NKT, 10 W)~\cite{wilson2011}. Then, this light is converted to 302 nm in a resonant cavity, achieving approximately 50 mW output power.

The cavity is followed by an always-on AOM for amplitude stabilization, and a pulsed AOM to generate the gate pulses (Fig. \ref{fig:uv_laser}a). The pulsed light is coupled into a solarization-resistant UV fiber patchcord~\cite{colombe2014} (NKT LMA-PM-15) and delivered to a monolithic breadboard mounted directly next to the glass cell, where it is focused to a beam waist of $w_0 = 10\,\mu$m at the atoms using an objective lens (Thorlabs LMU-3X-UVB). The power on the breadboard is approximately 6 mW. The entire breadboard is mounted on a motorized stage to align the beam to the atoms. The free-space optical path length after the fiber is less than 30 cm, reducing sensitivity to air currents and temperature gradients. A photodiode on the breadboard monitors the pulse power. Additionally, approximately 10\% of the light is picked off and coupled back into a second UV fiber and beat against the un-modulated UV laser in a heterodyne configuration. This allows the complex envelope of the laser pulse to be measured, to adjust the driving signal to compensate for phase transients during the rising and falling edges (Fig.~\ref{fig:uv_laser}b,c).

\subsection{Characterization of state preparation and measurement errors}

While we initialize and measure atoms in the tweezers with a fidelity of approximately 0.995, the qubit initialization and measurement of the qubit are affected by additional errors. These errors are dominated by the loss of atoms, including loss during the optical pumping into and out of $^3P_0$ (we believe the return step is the dominant source of loss) and decay out of $^3P_0$ during the gate sequence. 

With the destructive spin readout scheme used in this work, qubits in $\ket{1}$ cannot be distinguished from atom loss. However, our non-destructive imaging errors are biased towards false negatives: from repeated imaging, we infer a false positive atom detection probability of $4 \times 10^{-4}$. Therefore, the SPAM errors predominately result in incorrect assignment of an atom in bright state to dark due to loss, while the probability of assigning a lost atom to bright is very low. Exploiting this bias, we devise an accurate method to correct for the atom losses by converting all observable to the probability of both atoms being bright $P_\mathrm{bb}$. We first explain the theory of atom loss correction by measuring $P_\mathrm{bb}$, and then show how to convert Bell state fidelity to $P_\mathrm{bb}$. 

For a generic process, $P_\mathrm{bb}$ can be written as 
\begin{equation}
P_\mathrm{bb}=P_\mathrm{(bb|nl)}P_\mathrm{nl}+P_\mathrm{(bb|loss)} P_\mathrm{loss},
\end{equation}
where $P_\mathrm{nl}$ is the probability of no atom loss during the process, $P_\mathrm{loss}$ is the probability of at least one atom lost, $P_\mathrm{(bb|nl)}$ and $P_\mathrm{(bb|loss)}$ are the corresponding conditional probability of getting both atoms in the bright state. From the above discussion, the probability that a lost atom appears bright is negligible, $P_\mathrm{(bb|loss)}\approx 0$, so $P_\mathrm{bb}=P_\mathrm{(bb|nl)} P_\mathrm{nl}$. $P_\mathrm{nl}$ can be measured independently by removing the CZ gate and blowout pulse. If the spin readout is perfect, then the conditional fidelity $P_\mathrm{(bb|nl)} = P_\mathrm{bb}/P_\mathrm{nl}$ gives the fidelity corrected for atom loss during the initialization and readout process. For completeness, we assume the spin readout is perfect for now and provide a lower bound on the Bell state fidelity after introducing the detailed experimental scheme.

Here we explain in detail how we apply this correction to the Bell state fidelity in Fig. 2 in the main text. The Bell state fidelity can be determined as $\mathcal{F}_{B}=(P_{00}+P_{11}+P_c)/2$, so we need to convert the three terms on the right-hand side to $P_\mathrm{bb}$. Since $\ket{0}$ is the bright state, $P_{00}$ can be directly measured as $P_\mathrm{bb}$. $P_{11}$ can be measured by applying an additional nuclear spin $\pi$ pulse and then measuring $P_\mathrm{bb}$.

To measure $P_c$, we need an observable $\mathcal{O}_c=\ket{11}\bra{00}+\ket{00}\bra{11}$, which can be obtained using the parity oscillation circuit in Fig. 3c. This circuit can be represented as $U = e^{-i\pi X/4}e^{-i\theta Z/2}$, where $X= \sigma_x^1+\sigma_x^2$, $Z=\sigma_z^1+\sigma_z^2$, with $\sigma_x^j$, $\sigma_z^j$ being the Pauli operator on $j^\mathrm{th}$ atom. The probability to measure $P_\mathrm{bb}$ is then $P_\mathrm{bb} = \mathrm{Tr}(U \rho U^\dagger \mathcal{O}_\mathrm{bb})$, with  $\mathcal{O}_\mathrm{bb}=\ket{00}\bra{00}$.

This is mathematically equivalent to measuring an effective observable 
\begin{equation}
\begin{aligned}
\mathcal{O}_\theta &=e^{i\theta Z/2}e^{i\pi X/4} \mathcal{O}_{00} e^{-i\pi X/4}e^{-i\theta Z/2}\\
& =\frac{1}{4}
\begin{pmatrix}
    1 & -ie^{i\theta} & -ie^{i\theta} & -e^{2i\theta}\\
    ie^{-i\theta} & 1 & 1 & -ie^{i\theta}\\
    ie^{-i\theta} & 1 & 1 & -ie^{i\theta}\\
    -e^{-2i\theta} & ie^{-i\theta} & ie^{-i\theta} & 1\\
\end{pmatrix}.
\end{aligned}
\end{equation}
The term oscillating at $2\theta$ is the desired observable, $\mathcal{O}_c/4$. By fitting the parity oscillation signal in Fig. 3e to $A\cos(2\theta+\theta_0) + B$, we obtain $P_c = 4A$.

In a second experiment, we measure the dimer survival probability $P_\mathrm{nl}$ without the CZ gate and spin blowout pulses. The intrinsic Bell state fidelity is estimated to be $\mathcal{F}_B^{c}=\mathcal{F}_B/P_\mathrm{nl}$. In the experiment, we measure $P_{00}=0.46(1), P_{11}=0.42(1), P_{c}=0.86(2)$ and $P_\mathrm{nl}=0.872(6)$, yielding $\mathcal{F}_{B} = 0.866(12)$ and a corrected value $\mathcal{F}_B^c=0.99(2)$.

So far we have assumed a perfect spin readout. Now we analyze the effect of spin readout infidelity, and derive an approximate lower bound on the intrinsic Bell state fidelity $\mathcal{F}_B^c$ in the presence of spin readout errors. In the following we discuss only the probabilities conditioned on no atom loss, and we drop the superscript $c$ for simplicity. We define the single-atom spin readout true positive rate $p_\mathrm{TP}$ (an atom in $\ket{0}$ appears bright), and false positive rate $p_\mathrm{FP}$ (an atom in $\ket{1}$ appears bright). Given an underlying, true Bell state population $P_{ij}$, the measured population $\tilde{P}_{ij}$ can be written as:
\begin{equation}
\begin{aligned}
\tilde{P}_\mathrm{00}&=P_{00}p_\mathrm{TP}^2+(P_{01}+P_{10})p_\mathrm{TP}p_\mathrm{FP}+P_{11}p_\mathrm{FP}^2
\\
\tilde{P}_\mathrm{11}&=P_{11}p_\mathrm{TP}^2+(P_{01}+P_{10})p_\mathrm{TP}p_\mathrm{FP}+P_{00}p_\mathrm{FP}^2,
\end{aligned}
\label{eqs:P00P11}
\end{equation}
where the $\tilde{P}_{11}$ is extracted by applying a $\pi$ pulse then measuring the double bright state population.
The diagonal part of the Bell state fidelity is then 
\begin{equation}
P_{00}+P_{11}=\frac{\tilde{P}_{00}+\tilde{P}_{11}-2p_\mathrm{TP}p_\mathrm{FP}}{(p_\mathrm{TP}-p_\mathrm{FP})^2}.
\end{equation}
To derive the coherence term, we rewrite Eq.~\ref{eqs:P00P11} using the observables with imperfect spin readout $\tilde{\mathcal{O}}_\mathrm{bb}=p_\mathrm{TP}^2\ket{00}\bra{00}+p_\mathrm{TP}p_\mathrm{FP}(\ket{11}\bra{00}+\ket{00}\bra{11})+p_\mathrm{FP}^2\ket{11}\bra{11}$. We similarly define $\tilde{\mathcal{O}}_{\theta}=e^{i\theta Z/2}e^{i\pi X/4} \tilde{\mathcal{O}}_\mathrm{bb} e^{-i\pi X/4}e^{-i\theta Z/2}$. The experimental signal $\mathrm{Tr}(\rho \tilde{\mathcal{O}}_{\theta})$ contains a $\cos{(2\theta+\theta_0)}$ oscillation term with amplitude $1/[4(p_\mathrm{TP}-p_\mathrm{FP})^2]$, therefore the measured coherence is $\tilde{P}_c=P_c (p_\mathrm{TP}-p_\mathrm{FP})^2$ with $P_c$ the actual coherence. Therefore, actual Bell state fidelity $\mathcal{F}_B$ is related to the measured fidelity $\tilde{\mathcal{F}}_B$ via
\begin{equation}
\begin{aligned}
\mathcal{F}_B&=\frac{\tilde{\mathcal{F}}_B-p_\mathrm{TP}p_\mathrm{FP}}{(p_\mathrm{TP}-p_\mathrm{FP})^2}\\
&\ge \frac{\tilde{\mathcal{F}}_B-p_\mathrm{FP}}{(1-p_\mathrm{FP})^2}\\
& = \tilde{\mathcal{F}}_B + (2\tilde{\mathcal{F}}_B-1)p_\mathrm{FP}+(3\tilde{\mathcal{F}}_B-2)p_\mathrm{FP}^2 + O(p_\mathrm{FP}^3) \\
& \ge \tilde{\mathcal{F}}_B,
\end{aligned}
\end{equation}
where the inequality in the second line is obtained by setting $p_\mathrm{TP}=1$ (the inequality holds for $p_\mathrm{TP}\ge p_\mathrm{FP}$), and the inequality in the last line holds as long as $\tilde{\mathcal{F}}_B$ is significantly greater than $1/2$ and $p_{FP}$ is small. In our case, $p_\mathrm{FP}=0.4\%$ and $\tilde{\mathcal{F}}_B=0.99(2)$, so the measured value is a lower bound on the true value without spin readout errors.

\subsection{Two-qubit entangling gates}

\begin{figure*}
    \centering
    \includegraphics[width=180mm]{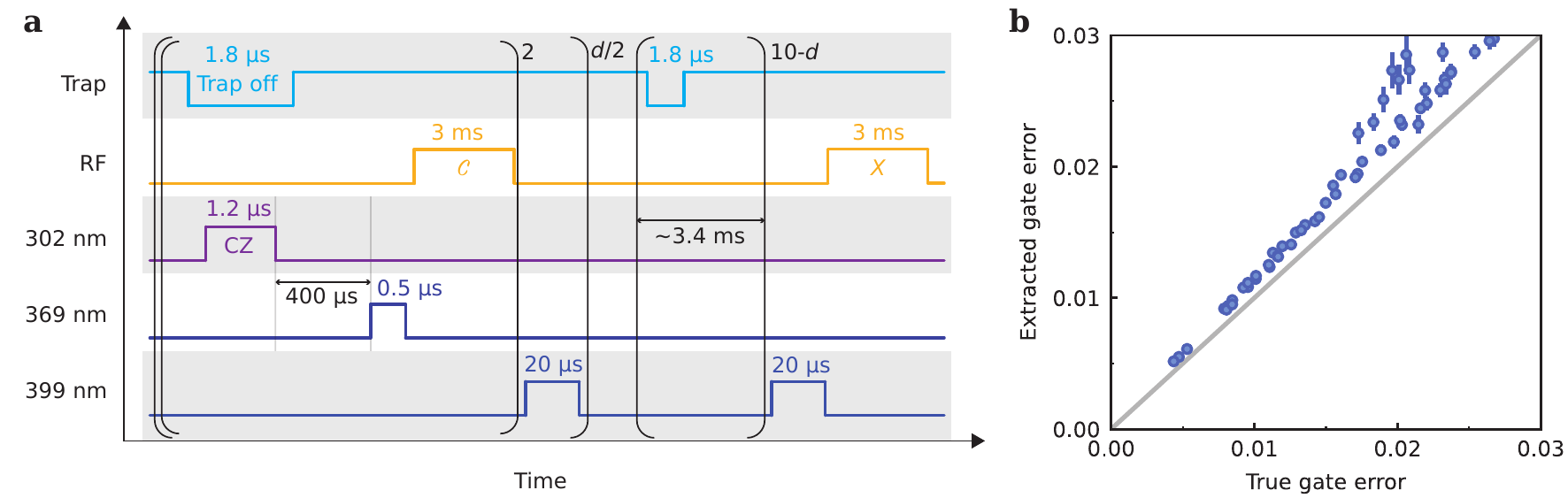}
    \caption{\textbf{Randomized circuit benchmarking experiment} (a) Sequence of operations for the entangling gate randomized circuit benchmarking experiment (the horizontal axis is not to scale). (b) Comparison of the simulated gate fidelity and simulated randomized circuit benchmarking fidelity for the error model discussed in the Methods. The strength of each noise term is varied randomly around the nominal value expected for the experiment. In all cases, the randomized circuit benchmarking infidelity is higher than the true gate error, supporting the relevance of this benchmark as a lower bound on the true gate fidelity.}
    \label{fig:method_srb}
\end{figure*}

The two-qubit gate implemented in our experiment is adapted from the time-optimal gate in Ref.~\cite{jandura2022}. In that work, each atom is modeled as a three-level system $\left\{\ket{0}, \ket{1}, \ket{r}\right\}$ with a perfect Rydberg blockade preventing a simultaneous excitation of both atoms to $\ket{r}$, and with a coupling of $\ket{1}$ and $\ket{r}$ through a global laser with constant amplitude and  time-dependent phase $\phi(t)$. Using the quantum optimal control method of Gradient Ascent Pulse Engineering (GRAPE) \cite{khaneja2005, wilhelm2020}, Ref. \cite{jandura2022} then determines the time-optimal pulse $\phi(t)$ to implement a CZ gate.

This simple three-level model does not accurately describe our system, because of off-resonant coupling between both qubit states and other Rydberg levels (Fig.~\ref{fig:3P0_init_readout}d). To incorporate this effect, we use GRAPE to redesign our pulses under a new model that takes all these additional transitions into consideration. All four sublevels of the $\ket{6s59s,\,^3S_1,\,F=3/2}$ Rydberg manifold are included: $\{\ket{r_{-3/2}}, \ket{r_{-1/2}}, \ket{r_{1/2}}, \ket{r_{3/2}}\}$. Taking the polarization of our Rydberg laser and the Clebsch-Gordon coefficients of each transition into account, the Hamiltonian of a single atom in the basis of $\{\ket{0}, \ket{1}, \ket{r_{-3/2}}, \ket{r_{-1/2}}, \ket{r_{1/2}}, \ket{r_{3/2}}\}$ can then be written as follows:
\begin{equation}
    H_{sq} = \hbar
        \begin{pmatrix}
        -\Delta_m & 0 & \frac{\Omega e^{-i\phi}}{2} & 0 & \frac{\Omega e^{-i\phi}}{2\sqrt{3}} & 0 \\
        0 & 0 & 0 & \frac{\Omega e^{-i\phi}}{2\sqrt{3}} & 0 & \frac{\Omega e^{-i\phi}}{2} \\
        \frac{\Omega e^{i\phi}}{2} & 0 & -3\Delta_r & 0 & 0 & 0 \\
        0 & \frac{\Omega e^{i\phi}}{2\sqrt{3}} & 0 & -2\Delta_r & 0 & 0 \\
        \frac{\Omega e^{i\phi}}{2\sqrt{3}} & 0 & 0 & 0 & -\Delta_r & 0  \\
        0 & \frac{\Omega e^{i\phi}}{2} & 0 & 0 & 0 & 0 \\
    \end{pmatrix}.
\end{equation}

Here, $\Omega$ is the Rabi frequency, and $\phi$ is the  phase of the Rydberg laser. The Zeeman splitting in the $^3P_0$ and Rydberg manifolds is denoted by $\Delta_m$ and $\Delta_r$, respectively. Because the Land\'e g-factor in the $^3P_0$ manifold is more than three orders of magnitude smaller than the one in $6s59s~^3S_1~F=3/2$, we set $\Delta_m = 0$ for simplicity. When taking both atoms and the van der Waals interaction into consideration, the full Hamiltonian of the system is then
\begin{equation}
    H = H_{sq} \otimes \mathds{I} + \mathds{I} \otimes H_{sq} + \hbar \sum_{ijkl} V_{ijkl} \ket{r_i}\bra{r_k} \otimes \ket{r_j}\bra{r_l}
\end{equation}
In the limit of strong van der Waals interaction ($|V_{ijkl}|\gg \Omega$), any double-Rydberg excitation is strictly forbidden. In this case, one can separately calculate the dynamics of the system depending on its initial state, with each corresponding to an evolution in a five-dimensional subspace:
\begin{align*}
\ket{00} & \Rightarrow \left\{\ket{00}, \ket{0r_{-3/2}}, \ket{0r_{1/2}}, \ket{r_{-3/2}0}, \ket{r_{1/2}0}\right\}, \\
\ket{01} & \Rightarrow \left\{\ket{01}, \ket{0r_{-1/2}}, \ket{0r_{3/2}}, \ket{r_{-3/2}1}, \ket{r_{1/2}1}\right\}, \\
\ket{11} & \Rightarrow \left\{\ket{11}, \ket{1r_{-1/2}}, \ket{1r_{3/2}}, \ket{r_{-1/2}1}, \ket{r_{3/2}1}\right\}.
\end{align*}

We note that the dynamics of $\ket{01}$ and $\ket{10}$ are always the same, and therefore the latter is omitted for brevity.

Given a specific value of $\Delta_r/\Omega$, a GRAPE optimization similar to Ref.~\cite{jandura2022} can be implemented with our more accurate model. Instead of using a pulse with square amplitude, we fix $\Omega(t)$ to have Gaussian rising and falling edges, and total duration $T$. This minimizes the pulse bandwidth and reduces unwanted excitation of the other Rydberg state, while having negligible effect on the average population of the Rydberg state. We then find the laser phase $\phi(t)$ minimizing the infidelity for a CZ gate. For the sake of the optimization, a piecewise constant approximation $\phi(t) = \phi_n$ for $t \in [Tn/N, T(n+1)/N]$  with $N=100 \gg 1$ pieces are made. The infidelity $1-\mathcal{F}$ can then be numerically minimized over the $\phi_0,...,\phi_{N-1}$, with the GRAPE algorithm providing an efficient way to calculate the gradient $\nabla \mathcal{F}$ \cite{khaneja2005} in time $\mathcal{O}(N)$. Note that also a global phase $\theta_0$ and a single qubit phase $\theta_1$ are included in the optimization, such that the desired evolution is given by $\ket{00} \mapsto e^{i\theta_0}\ket{00}$, $\ket{01} \mapsto e^{i(\theta_0+\theta_1)} \ket{01}$ and $\ket{11} \mapsto - e^{i(\theta_0+2\theta_1)}\ket{11}$. 

For the experimental parameters of $\Delta_r = 2\pi\times 9.3$ MHz and $\Omega = 2\pi \times 1.6$ MHz,  $\Delta_r/\Omega = 5.8$. For these parameters, we find a gate with infidelity $1- \mathcal{F} < 10^{-5}$.

For deployment on the experiment, we use a parameterized version of the GRAPE-derived pulse in terms of a finite sum of Chebyshev polynomials:
\begin{equation}
    \Delta(t) = \dot\phi (t) \approx \sum_{n=0}^{n_\mathrm{max}} c_n T_n\left(\frac{2t}{T} - 1\right),
\end{equation}
where $T_n(x)$ is the $n$-th Chebyshev polynomial of the first kind. We find that truncating the series at $n_\mathrm{max} = 13$ does not impact the gate fidelity. This lower dimensional parameterization is useful for experimental fine-tuning by scanning each coefficient  around its nominal value, to correct for control errors and effects not included in our model.

\subsection{Noise model for two-qubit gates}

To understand the sources of infidelities in our two-qubit gates, we have developed a numerical simulation combining the master equation formalism with the Monte Carlo method, based on the six-level model discussed above. It includes Markovian decay from the finite Rydberg state lifetime, coherent errors (imperfect Rydberg blockade and off-resonant excitation), and non-Markovian noise (Doppler shifts from atomic motion, and laser phase and intensity noise). Non-Markovian effects are included using a Monte Carlo approach, by simulating the evolution under randomly generated noise traces and averaging the final result~\cite{deleseleuc2018a}.

The parameters in the error model are determined from independent experiments. Exploiting the ability to trap Rydberg atoms~\cite{wilson2022}, we directly measure the Rydberg state lifetime, finding $T_{1,r} = 65(2)\,\mu$s. We use a Ramsey experiment to measure the Doppler shift and other quasi-static detuning errors and find a purely Gaussian decay with $1/e$ decay time $T_2^* = 5.7\,\mu$s. This places an upper bound on the temperature of $T \leq 2.94\,\mu$K. We measure the laser phase noise before the second harmonic generation, using a high-finesse cavity as a frequency discriminator. Finally, we measure the laser intensity noise after the second harmonic generation.

We find that the leading sources of error are the Rydberg state decay ($4 \times 10^{-3}$), detuning from Doppler shifts ($5 \times 10^{-3}$), laser phase noise ($2 \times 10^{-3}$), and imperfections in the laser pulse envelope ($2 \times 10^{-3}$). Simulating these effects together gives a gate error of $1.1 \times 10^{-2}$, lower than the experimental value of $2 \times 10^{-2}$. We attribute some of this discrepancy to the fact that the randomized circuit benchmarking systematically overestimates the error (see next section). However, this leaves an error of approximately $5 \times 10^{-3}$ that is not accounted for in our model. The most plausible explanation is slow drifts in experimental parameters affecting the gate calibration, but further investigation is required to isolate and correct this error.

The model additionally predicts that 60\% of the errors should be leakage errors, including decays from the Rydberg state during the gate and population trapped in the Rydberg state at the end of the gate. This is consistent with the observed erasure conversion fraction of 33\% for the two-qubit gates (Fig. 4c) when correcting for the fact that we only detect 50\% of the decays from the Rydberg state in $^1S_0$.

To achieve higher gate fidelities, the key parameter is the Rabi frequency. The Rydberg decay error decreases with the gate time, as $1/\Omega_{UV}$. The Doppler shift error decreases as $1/\Omega_{UV}^2$. Given that our current UV laser power of 6 mW is far from the highest power demonstrated at a similar wavelength~\cite{lo2014}, the known error sources can all be suppressed below $10^{-3}$ with straightforward improvements in laser power, phase noise and pulse shape control.

\subsection{Randomized circuit benchmarking validation}

As noted in the text, the two-qubit benchmarking circuit used in Fig. 3f and Fig. 4c does not generate a rigorous fidelity estimate because we use global single-qubit rotations. It is completely insensitive to certain errors, such as a SWAP of the two qubits~\cite{baldwin2020}. To assess the reliability of this estimate, we have simulated the exact benchmarking sequences used in the experiment with the error model discussed in the previous section, varying the strength of each noise term over a significant range to simulate a range of gate error rates, up to several times worse than the measured experimental value. In Fig.~\ref{fig:method_srb}b, we compare the true gate error rates with the those extracted from the simulated benchmarking circuit, and find that the benchmarking estimate consistently overestimates the error rate, providing a lower bound on the gate fidelity.

\bibliography{jdt_erasure}

\end{document}